\documentclass[sigconf,10pt]{acmart}
\setcopyright{none}
\renewcommand\footnotetextcopyrightpermission[1]{}
\acmDOI{}

\usepackage{balance}
\usepackage{amsmath}
\usepackage{algorithmic}
\usepackage{graphicx}
\usepackage{textcomp}
\usepackage{xcolor}
\usepackage{multirow}
\usepackage{csquotes}
\usepackage{comment}
\usepackage{float}
\usepackage{afterpage}

\hypersetup{breaklinks=true}
\def\BibTeX{{\rm B\kern-.05em{\sc i\kern-.025em b}\kern-.08emT\kern-.1667em\lower.7ex\hbox{E}\kern-.125emX}}
\begin{document}

\title{An Experimental Analysis on Drone-Mounted Access Points for Improved Latency-Reliability}

\author{Igor Donevski}
\affiliation{%
  \institution{Aalborg University}
  \city{Aalborg} 
   \country{Denmark} 
}
\email{igordonevski@es.aau.dk}

\author{Christian Raffelsberger}
\affiliation{%
  \institution{Lakeside Labs GmbH}
  \city{Klagenfurt} 
   \country{Austria} 
}
\email{raffelsberger@lakeside-labs.com}

\author{ Micha Sende}
\affiliation{%
  \institution{Lakeside Labs GmbH}
  \city{Klagenfurt} 
   \country{Austria} 
}
\email{sende@lakeside-labs.com}

\author{Aymen Fakhreddine}
\affiliation{%
  \institution{Lakeside Labs GmbH}
  \city{Klagenfurt} 
   \country{Austria} 
}
\email{fakhreddine@lakeside-labs.com}

\author{Jimmy Jessen Nielsen}
\affiliation{%
  \institution{Aalborg University}
  \city{Aalborg} 
   \country{Denmark} 
}
\email{jjn@es.aau.dk}

\begin{abstract}
The anticipated densification of contemporary communications infrastructure expects the use of drone small cells (DSCs). Thus, we experimentally evaluate the capability of providing local and personalized coverage with a drone mounted Wi-Fi access point that uses the nearby LTE infrastructure as a backhaul in areas with mixed line of sight (LoS) and Non-LoS (NLoS) links to the local cellular infrastructure. To assess the potential of DSCs for reliable and low latency communication of outdoor users, we measure the channel quality and the total round trip latency of the system. For a drone following the ground user, the DSC-provided network extends the coverage for an extra 6.4\% when compared to the classical LTE-direct link. Moreover, the DSC setup provides latencies that are consistently smaller than 50\,ms for 95\% of the experiment. Within the coverage of the LTE-direct connection, we observed a latency ceiling of 120\,ms for 95\% reliability of the LTE-direct connection. The highest latency observed for the DSC system was 1200\,ms, while the LTE-direct link never exceeded 500\,ms. As such, DSC setups are not only essential in NLoS situations, but consistently improve the latency of users in outdoor scenarios.
\end{abstract}




\maketitle
\sloppy
\pagestyle{empty}
\section{Introduction}
Unmanned Aerial Vehicles (UAVs) -- or drones -- are considered a prime candidate for custom solutions to wireless communication coverage in both conventional and unconventional circumstances. Mainly as a benefit to their mobility and their altitude, they are expected to offer good wireless channel conditions towards ground users in outdoor scenarios  \cite{mainsurvey,tutorial}. Commonly referred to as Drone Small Cells (DSCs) or as Drone provided Access Points, the flying devices that carry wireless equipment are mostly considered as a custom solution for improving regional communication quality \cite{igoreucnc}. However, it is common among DSC literature to lay strong assumptions for the availability of backhaul and/or the propagation setting of the fronthaul \cite{mainsurvey}. While this is important for modeling future drone-specific network requirements, we turn the attention towards assessing the current state of commercially available technologies that are capable of drone provided communication services.

In this work we aim to aid the cellular network by analyzing and measuring the impact on latency for a mobile user (UE), located on the ground and in the presence of a DSC that provides a Wi-Fi hotspot, instead of a pure LTE relay.
The reason behind designing our DSC like this is that LTE positioning requires careful spectral planning which is not an approachable method for casual users of the DSC technology. Hence, we split the backhaul as LTE provided and the fronthaul offered through a Wi-Fi interface. This is performed by placing the DSC at an altitude that allows the utilization of a nearby Base Station (BS). The goal of this setup is to answer three key questions: does a drone provided hotspot improve the latency of the application, what are the shortcomings of such a system, and is this technology sufficient for use in reliable and low latency communications? Answering these questions is critical for prospective use of DSCs in cases of ground remote control applications that need to be tolerant to faults of the local cellular coverage.

The paper is organized as follows: Section~\ref{sec:literature} discusses significant related experimental work with drones. 
Section~\ref{sec:preana} describes a route for the horizontal spatial coordinates that is bound to be covered by a mixture of strong LoS and NLoS links for the UE along the route. On this route we forgo two types of approaches, pretesting the LTE propagation conditions for a travelling node on the ground, or following the same route on a specific altitude. 
In Section~\ref{sec:messetup} we showcase the measurements of the impact on signal and delay when using the DSC network (UE-UAV-BS) and a singular (UE-BS) LTE link. %
Finally, in Section \ref{sec:conclusion} we draw several conclusions, state the significance of the provided data, and discuss open issues.

\section{Related Work}
\label{sec:literature}
There has been a significant interest in deriving experimental results for cellular connected drones. Mainly, the effect of interference carries a significance due to the likelihood of drones to establish a LoS with neighbouring cells, which would not be visible to ground UE. This causes extra interference, which adds extra complexity to the problem of BS handover, topics that are thoroughly covered in: \cite{kovacs2017interference,fakhreddine2019handover,nguyen2018ensure,van2016lte}. Moreover, the work of \cite{lin2019mobile} gives an elaborate overview of the common issues plaguing cellular connected drones. In terms of the performance of cellular connected drones, the work of \cite{hayat2019experimental} provides detailed experimental analysis of the throughput of drones connected to an LTE-A network.  Justifiably, due to rollout phase of 5G infrastructure the work of \cite{muzaffar2020first} experimentally evaluates the throughput performance of 5G connected drones.

As for UE-DSC connectivity, there is a strong experimental support of the LoS NLoS model of \cite{atg}, such as the work of \cite{qiu2017low}. The work of \cite{athanasiadou2019lte} analyzes the drone-BS connectivity of LTE for its potential use as backhaul in DSCs. However, the experimental works of \cite{li2016development} and \cite{gangula2018flying} are most in line with our goals of testing UE connectivity through a DSC. Moreover, the work of \cite{li2016development} is a trivial example of UAV-to-UAV communications as a relay system. While the work of \cite{gangula2018flying} is most in line with our goals for establishing a UE-DSC-BS relay, its contribution is though limited to a throughput investigation in a small scenario. In addition, all prior UE-UAV-BS connectivity work has majority of the attention focused on measuring signal strength and throughput. Given the vastly different nature of cellular and Wi-Fi connectivity, and the added system complexity, we are interested in the latency and therefore delay quality for UE served by BSs versus DSCs as clarified in the following sections.

\section{Experimental Setup}
\label{sec:preana}

The experiments are performed in a relatively controlled setting in the 5G Playground Carinthia testbed in the outskirts of Klagenfurt, Austria. All flights and ground UE patrols were taken along the same identical horizontal path with a constant speed of 1~m/s. The LTE connectivity in the area is provided by a single BS mounted on top of a building, and was predetermined for the experimental analysis. 
The LTE BS supports a data rate of 150\,Mbit/s for each 20\,MHz channel supporting a modulation of 256\,QAM per carrier. While the BS supports up to two carriers, only one carrier is used during the experimentation.

\begin{figure}[t]
\centering
\includegraphics[width=1\columnwidth]{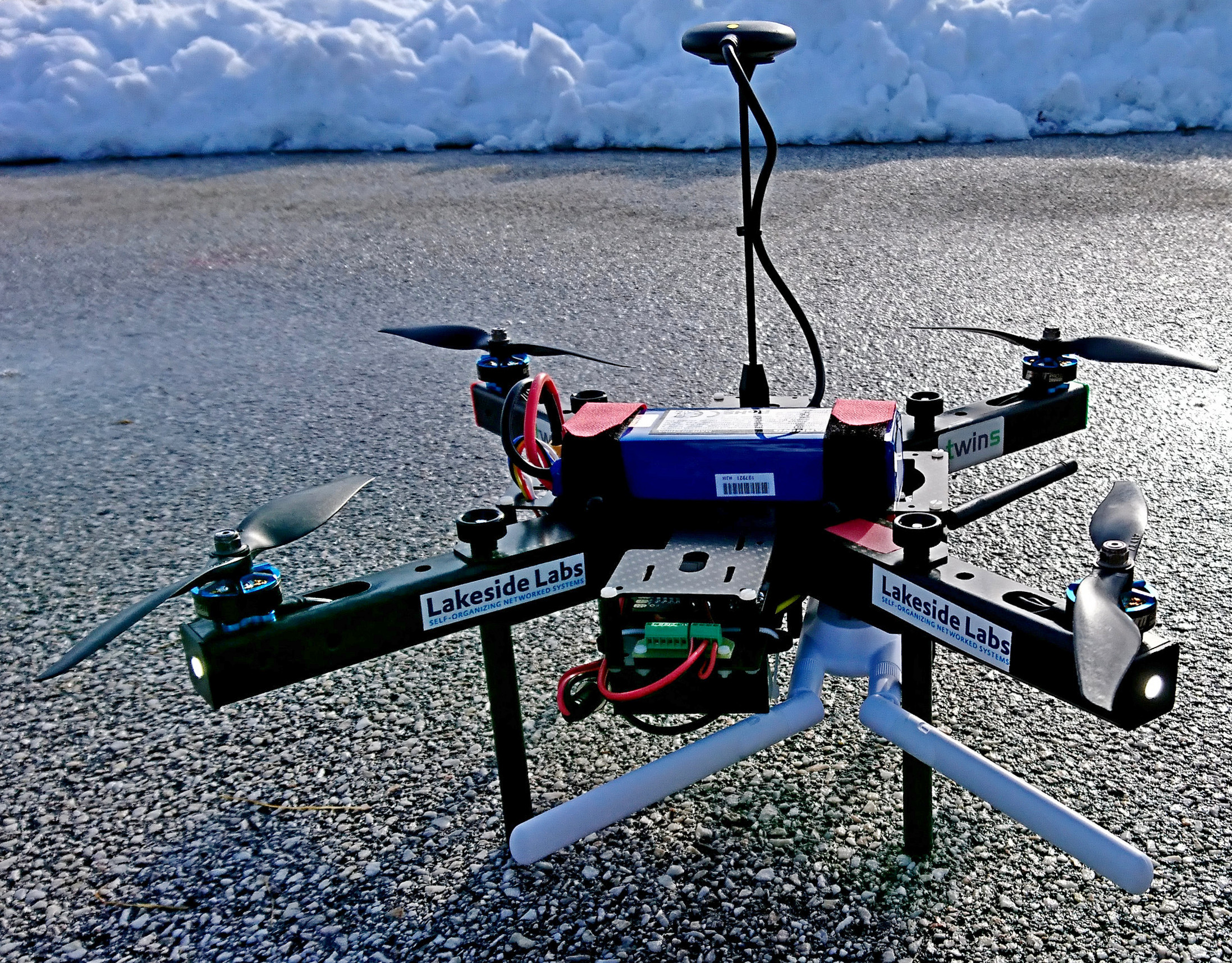}
\caption{An image taken from the DSC equipment mounted on a twinFOLD SCIENCE drone.
}
\label{fig:dronepic}
\end{figure}

The drone, Fig.~\ref{fig:dronepic}, is a twinFOLD SCIENCE by the Austrian vendor TWINS. The communications equipment mounted on the drone, consists of an LTE modem and a Wi-Fi 802.11ac access point connected to a Raspberry Pi 4 companion board that bridges the two interfaces and captures all traffic that passes through. This allows for a better investigation and separation of the fronthaul, backhaul and most importantly scheduling and processing delays. The Wi-Fi access point (Unifi UAP-AC-M) has a 2x2 WiFi dual band configuration with a maximum link data rate of 867 Mbit/s. The max. transmit power is 20\,dBm. The access point is connected to the main board via a  1\,Gbit/s Ethernet interface. The Huawei E3372 LTE modem supports LTE Cat4 with a max. downlink rate of 150\,Mbit/s and an uplink rate of 50\,Mbit/s. It is connected via USB to the companion board. The smartphone is a Samsung S20 5G which supports dual-band Wi-Fi 802.11ac with VHT80 MU-MIMO and LTE Cat20 with 4x4~MIMO.

The measurements are performed with the Cellular Drone Measurement Tool (CDMT) \cite{cdmt} that can record multiple parameters for the cellular connection such as:  reference signal received power (RSRP), reference signal received quality (RSRQ), received signal strength indication (RSSI), serving physical cell identity (PCI), and channel quality indicator (CQI) and throughput as averages of one second. To perform latency measurements, the CDMT app sends 10 UDP datagrams per second, containing a sequence number and two timestamp fields (20~byte payload), to a server located in the testbed network. The app logs the timestamp when the packet is sent and when it receives the reply from the server. The timestamp fields are used to store the time when a packet is received at the server and when it is sent back to the client in order to determine the processing time at the server.

\subsection{Scenario Investigation}
\begin{figure}[t]
\centering
\includegraphics[width=0.65\columnwidth]{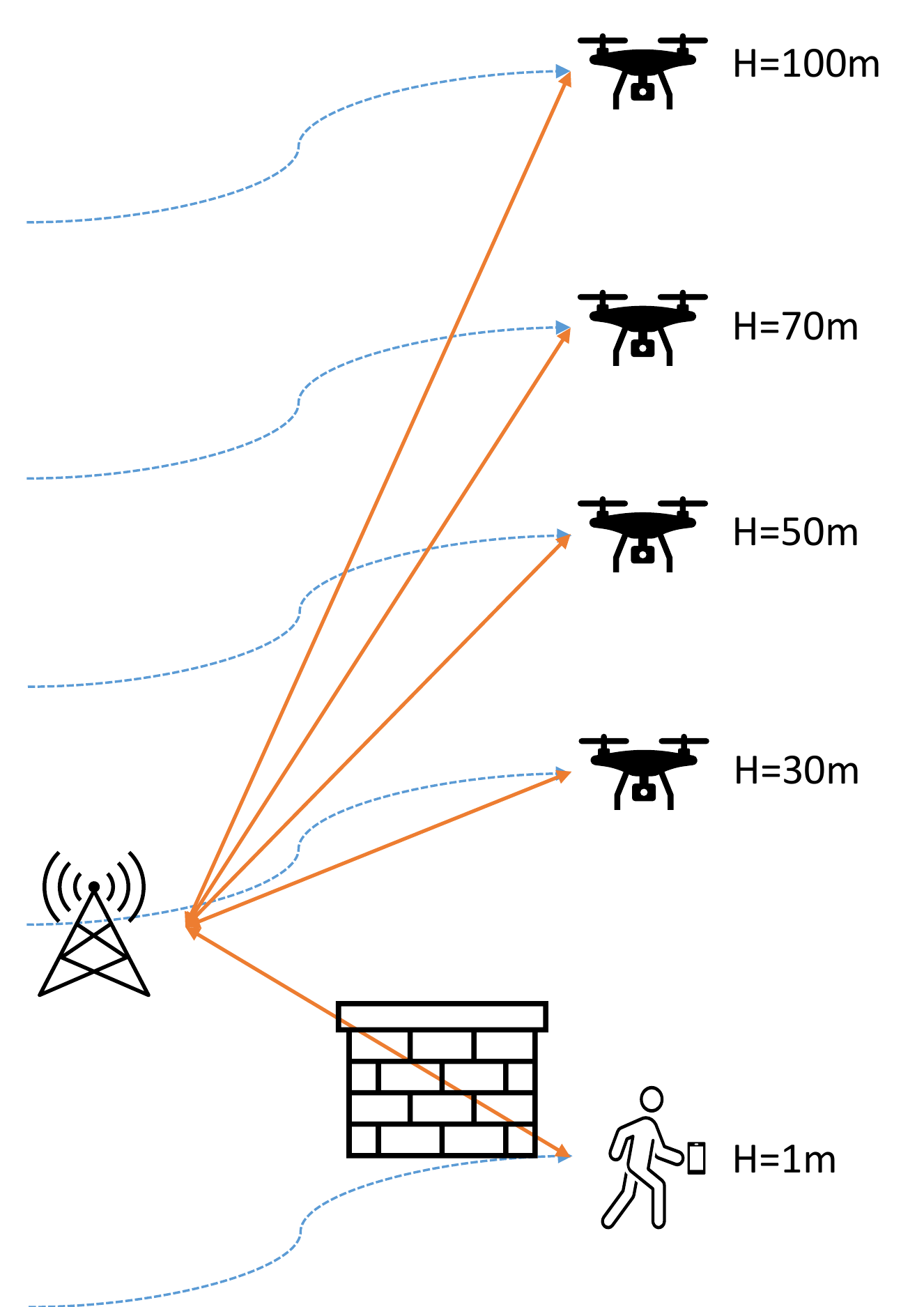}
\caption{Testing LTE signal performance for four altitudes (30\,m, 50\,m, 70\,m, 100\,m) and the ground UE-BS link is likely to be NLoS.
}
\label{fig:illupre}
\end{figure}
\begin{figure}[t]
\centering
\includegraphics[width=1\columnwidth]{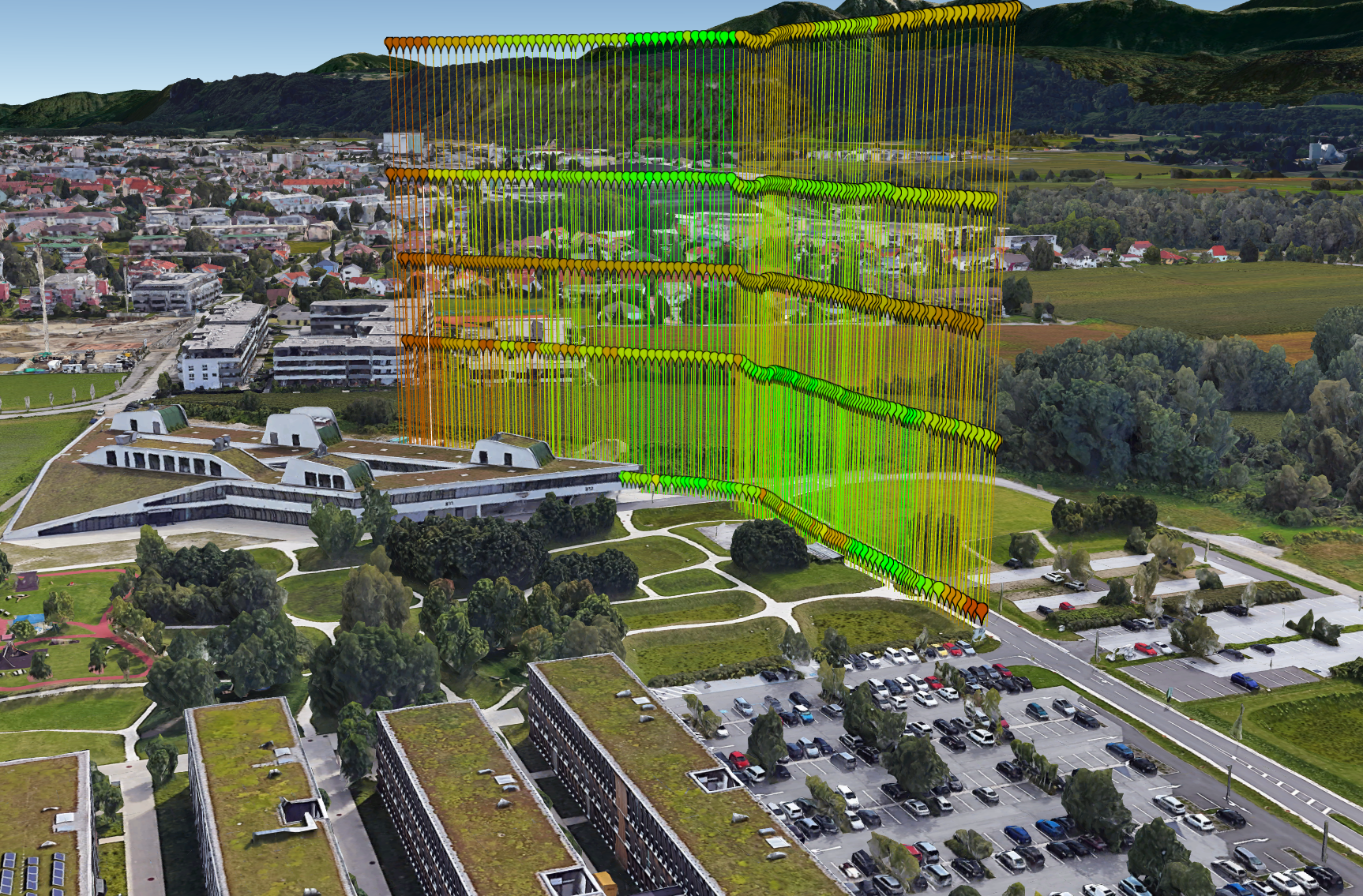}
\caption{RSSI measurement results for the LTE signal performance sweep from the perspective of the BS. Colormap ranges from red for the worst RSSI measured(-107 dB) to green for the best RSSI (-75 dB) measured. (Map data: Google ©2021)
}
\label{fig:premeasbs}
\end{figure}
The performance of the DSC is significantly influenced by the UAV-to-BS connectivity. It has been shown that the current cellular BS antenna design is optimized to service ground users and drones experience degraded performance at certain altitudes, since they are only covered by side lobes \cite{hayat2019experimental, muzaffar2020first}. To better understand the propagation pattern of the base station, we perform several pre-measurements in order to improve the deployment of the DSC (i.e., determine a suitable altitude for the drone). The pre-measurements start with measuring the performance on the ground, while a user carrying a smartphone walks along the route. Afterwards, we mount the smartphone on a UAV and repeat the same route at a different altitudes. Moreover, even when the drone can move horizontally its performance depends on how big of an area it wants to cover \cite{optlap}. Hence as per a previous analysis for suburban environments, we derive 30\,m as a decent altitude for a drone with an isotropic fronthaul (in our case Wi-Fi) transmitter, and 50\,m or 70\,m with a directional antenna \cite{igorwcnc}.  Furthermore, we also want to sample the possibility of having drone-to-BS connection with a drone at an altitude of 100\,m for UAVs that have an efficient directional antenna. As such we perform a sweep to test the LTE channel performance at different altitudes, as shown in Fig \ref{fig:illupre}.
\begin{figure}[t]
\centering
\includegraphics[width=1\columnwidth]{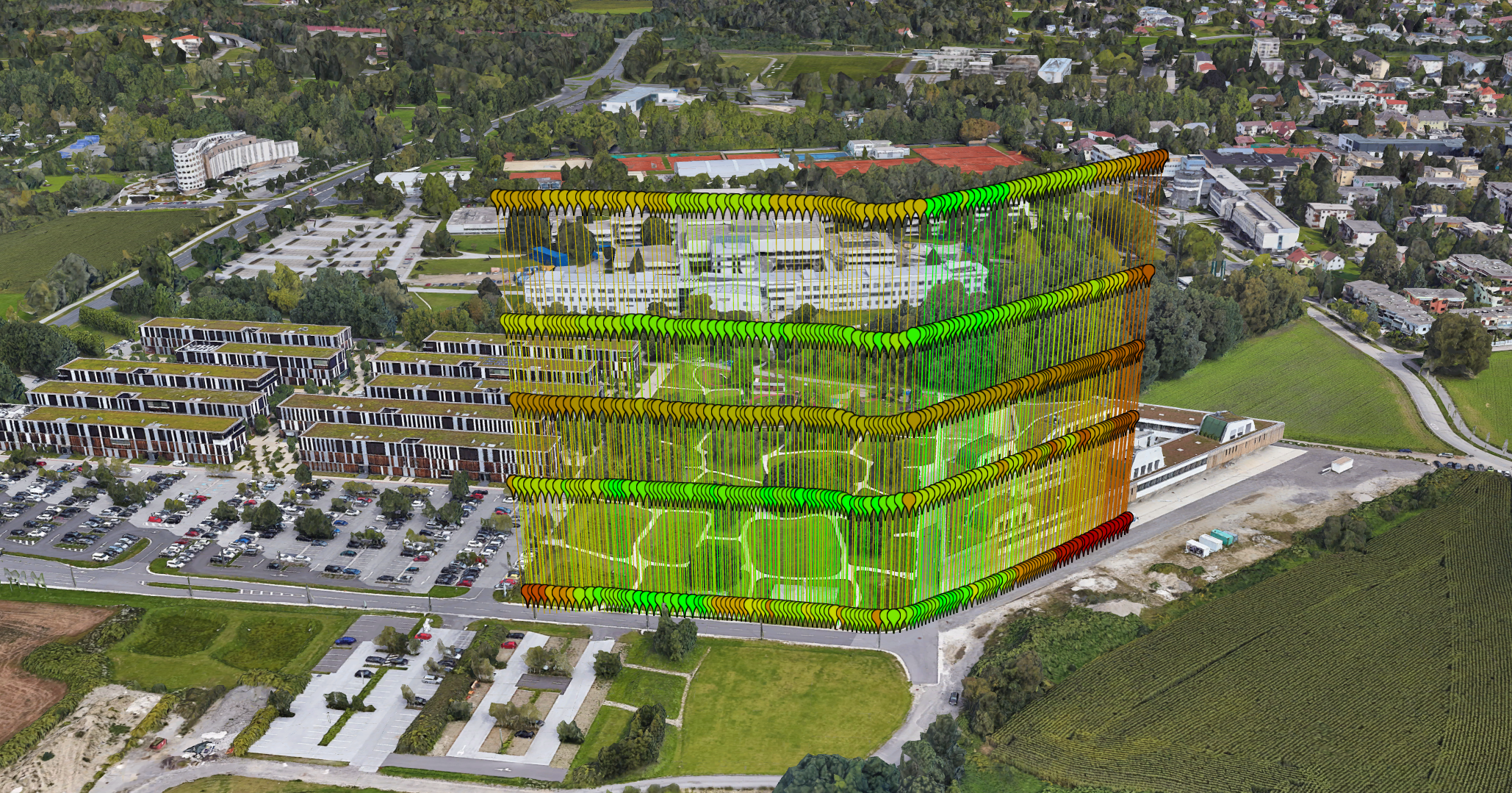}
\caption{RSSI measurement results for the LTE signal performance sweep from the perspective of the DSC. Colormap ranges from red for the worst RSSI measured(-107 dB) to green for the best RSSI (-75 dB) measured. (Map data: Google ©2021)
}
\label{fig:pers3}
\end{figure}

\begin{figure}[t]
\centering
\includegraphics[width=1\columnwidth]{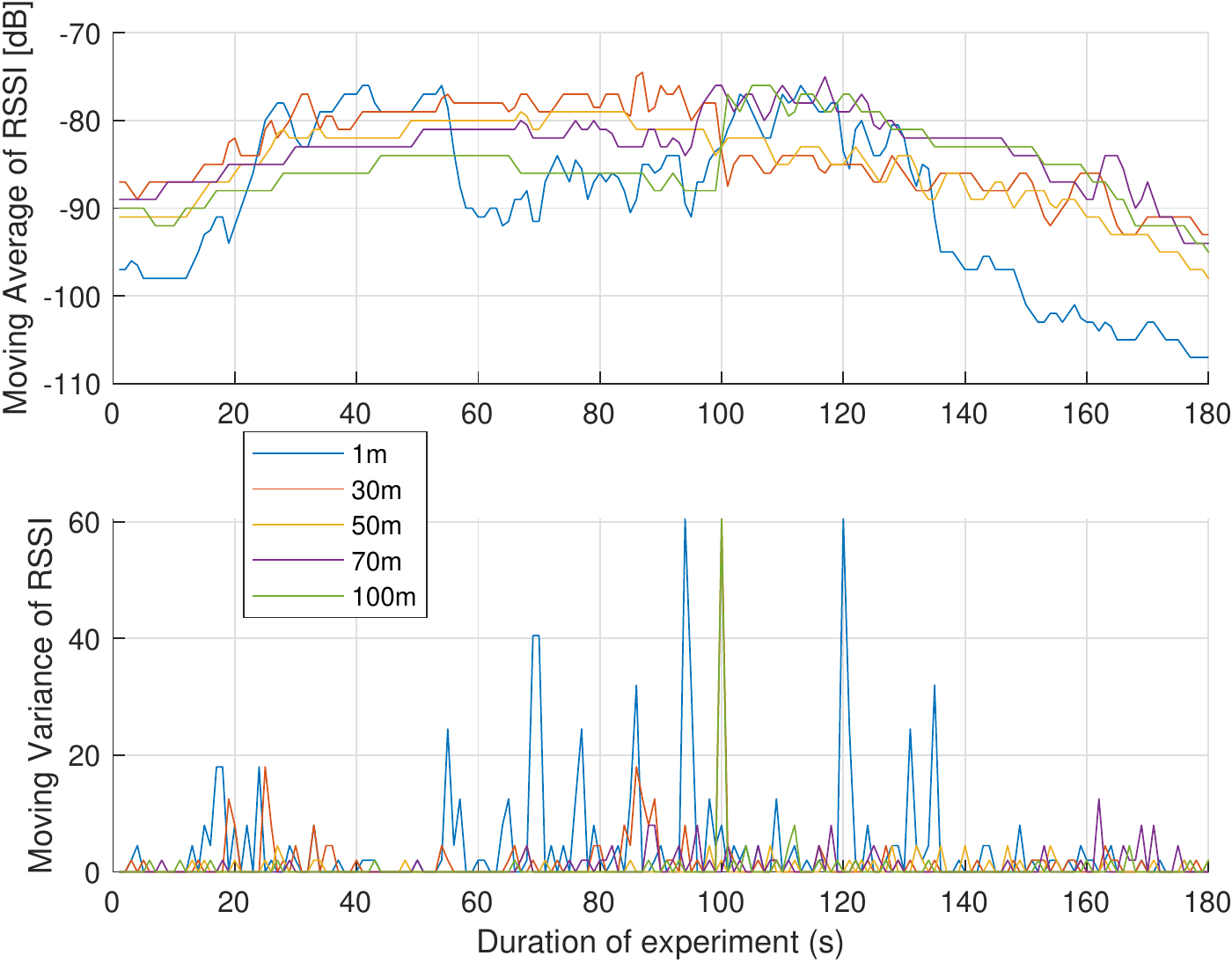}
\caption{2-point Moving Average and Moving Variance for RSSI values of all 5 experiments.
}
\label{fig:movs}
\end{figure}

In Fig. \ref{fig:premeasbs} we can observe that the altitudes of 30\,m and 70\,m give the best channel conditions. Since we conduct our experiments with omnidirectional Wi-Fi antennas, we select an altitude of 30\,m for the measurements of our DSC. Moreover, lower altitudes suffer less from  inter-cell interference and frequent handover challenges \cite{fakhreddine2019handover}. Looking at the RSSI measurements from the opposite side, as shown in Fig.~\ref{fig:pers3}, we can notice that the LTE connectivity of the ground UE severely deteriorates in the NLoS region. Thus, we expect that DSC-based connectivity should eliminate this issue and improve network performance for users hidden from the direct LoS of the BS.

A final remark of the pre-measurements can be derived from Fig.~\ref{fig:movs} where we show the moving evolution of variance and mean RSSI values. As it can be noticed, all flying drone implementations record lower RSSI fluctuations during the route. The signal strength at the moving ground UE was constantly changing, providing additional motivation for stabilizing the performance of the ground UE through a DSC system. The variance and the mean of the RSSI parameter for each pass of the sweep are contained in Table~\ref{table:RLmetrics}.
\begin{table}[t]
\begin{center}
\caption{Mean and variance of RSSI for all measured altitudes}
\begin{tabular}{llllll}
\hline
Label & 1\,m & 30\,m & 50\,m & 70\,m & 100\,m \\ \hline
\hline
mean $\mu$ & -89.79 & -84.46 & -85.98 & -84.06 & -86.83\\
variance $\sigma$ & 94.86 & 31.68 & 34.73 & 26.39 & 30.90\\
\end{tabular}
\label{table:RLmetrics}
\end{center}
\end{table}
\section{Experimental Results}
\label{sec:messetup}
\begin{figure}[t]
\centering
\includegraphics[width=1\columnwidth]{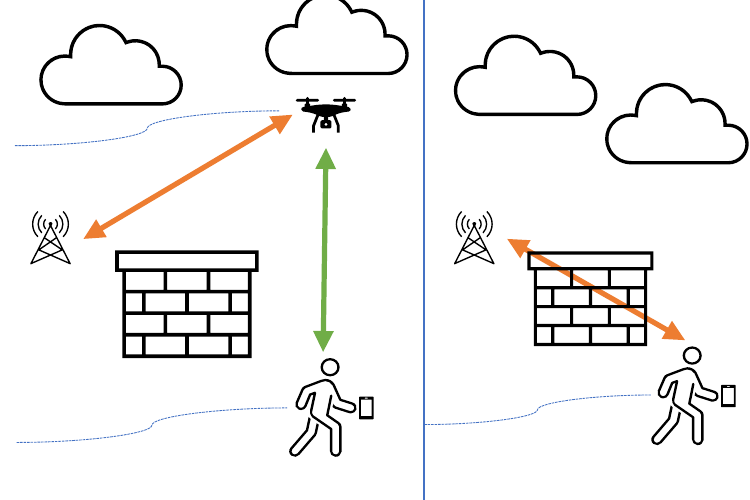}
\caption{Latency measurement scenarios: the drone and UE move at 1m/s together (left), the user moves at 1m/s alone (right)
}
\label{fig:messet}
\end{figure}

\begin{figure}[t]
\centering
\includegraphics[width=1\columnwidth]{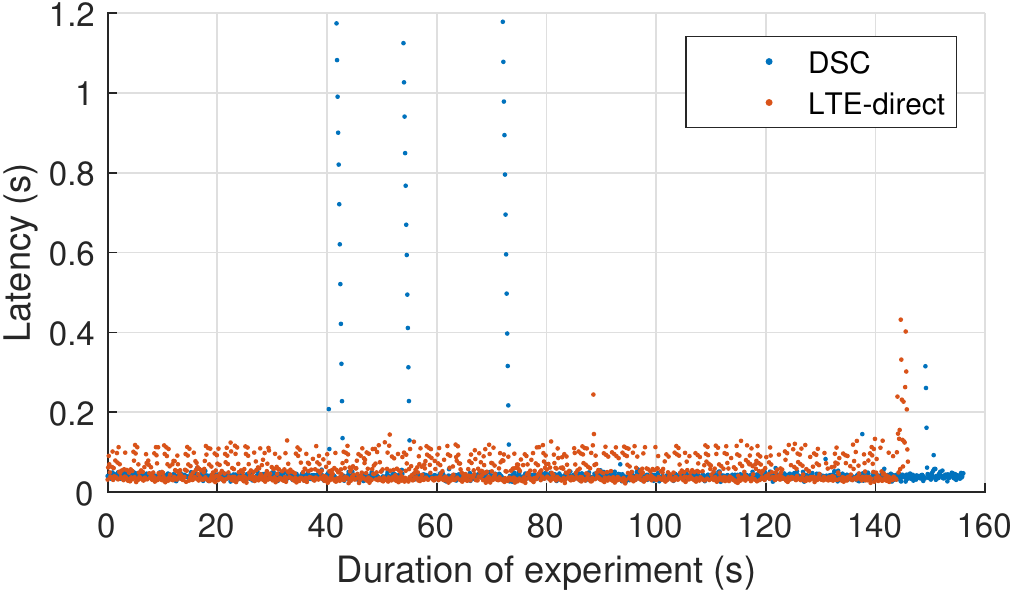}
\caption{The measured latency of the DSC and LTE-direct implementations.
}
\label{fig:m1}
\end{figure}

Since we have established the DSC altitude of 30\,m, we proceed with the latency measurements for the system at that particular altitude. As shown on the right in Fig.~\ref{fig:messet}, one part of the measurement is for the LTE-direct UE-BS link, measured along the route. On the left, in Fig.~\ref{fig:messet}, we illustrate the second part of the measurements where the Wi-Fi equipped drone provides DSC service to the mobile user, while flying directly above him or her along the same route. To verify the consistency of the measuring setup, the DSC measurements were repeated three times while the LTE-direct measurements were repeated two times. From that we observed consistent behavior across the measurement experiments. For better visibility, we illustrate the duration-time plot in Fig.~\ref{fig:m1} only for one sample of each type.

\begin{figure}[t]
\centering
\includegraphics[width=1\columnwidth]{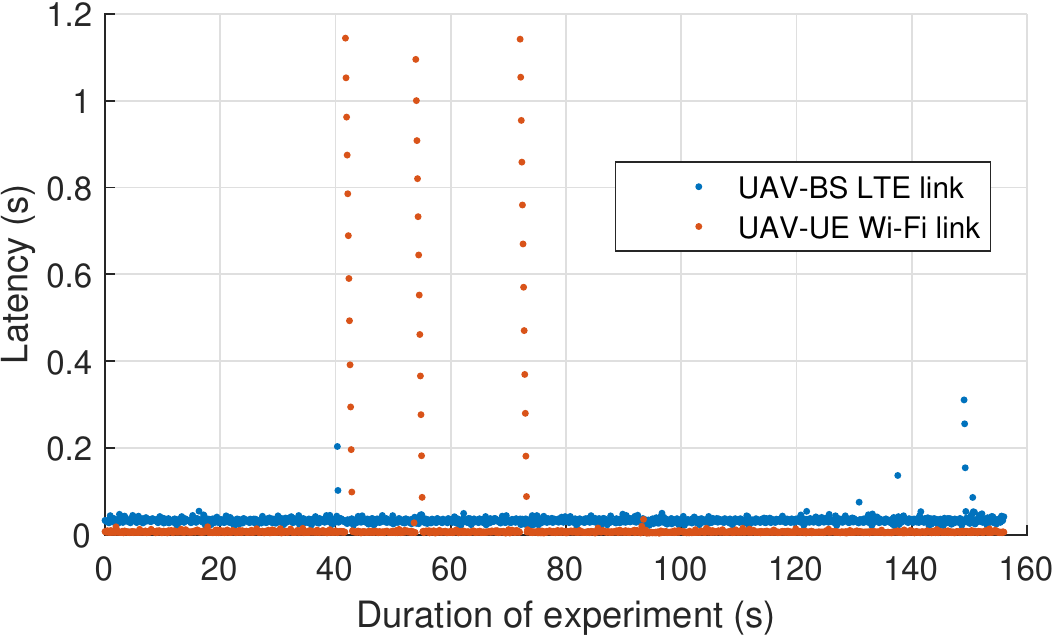}
\caption{Latency comparison of the Wi-Fi link versus the LTE link for the DSC setup.
}
\label{fig:m2}
\end{figure}

Since all experiments started from the same initial spot and travel with the same speed, in Fig.~\ref{fig:m1} we plot the latency evolution of the system with regards to the duration of the experiment. The most notable remark is that LTE-direct measurements lack 10\,s of data near the end of the experiment. This was due to the strong NLoS, shown with bright red in Fig.~\ref{fig:messet}, due to which the ground-UE consistently lost connectivity to the BS near the end of the experiment. This reduces the number of samples for the LTE-direct measurements by 6.4\% as for such portion of the time the LTE-direct link is in total outage. However, the DSC setup experiences latency spikes that are consistently above one second. These spikes occur in areas where the UAV has good LTE channel condition, with strong LoS. From the segmented latency data for the DSC measurement in Fig.~\ref{fig:m2}, we observe that the latency spikes are due to consistent and lengthy outages of the Wi-Fi interface. These spikes are unrelated to the position of the drone and occur at random times due to the distributed multiple access scheme implemented in Wi-Fi. Therefore, due to the inconsistency of Wi-Fi, the mean latency for both systems (within the area of BS coverage for ground users) is nearly identical, measuring to 54\,ms and 53.7\,ms of latency for the LTE-direct and the DSC connectivity, respectively. 
\begin{figure}[t]
\centering
\includegraphics[width=1\columnwidth]{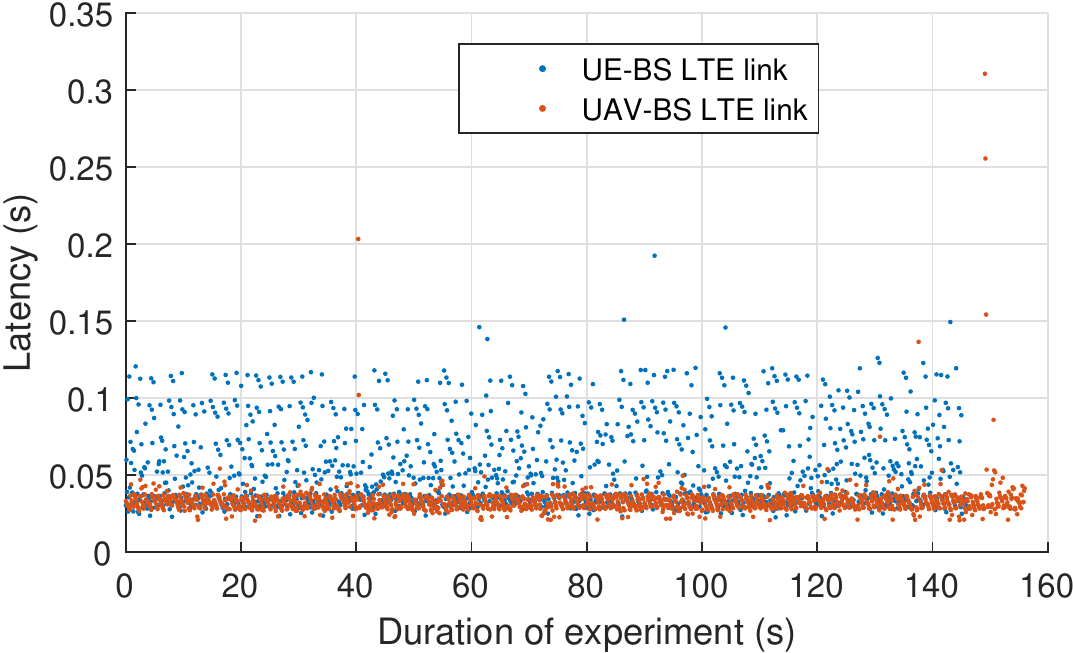}
\caption{Latency comparison of the LTE-direct links from the UE and the UAV/DSC.
}
\label{fig:m3}
\end{figure}
In addition, the Wi-Fi link adds insignificant amount of extra latency for the DSC case when it does not encounter crowding problems due to its distributed nature. Thus, in Fig.~\ref{fig:m3} we compare the latency between the UAV-BS LTE link to the UE-BS direct. The major superiority of the DSC system comes as a consequence of the good channel conditions of the flying drone, even when both setups have generally LoS signal. Moreover, the mean LTE delay from the DSC is 33.2\,ms while the mean LTE delay from the UE is 54\,ms.

\begin{figure}[t]
\centering
\includegraphics[width=1\columnwidth]{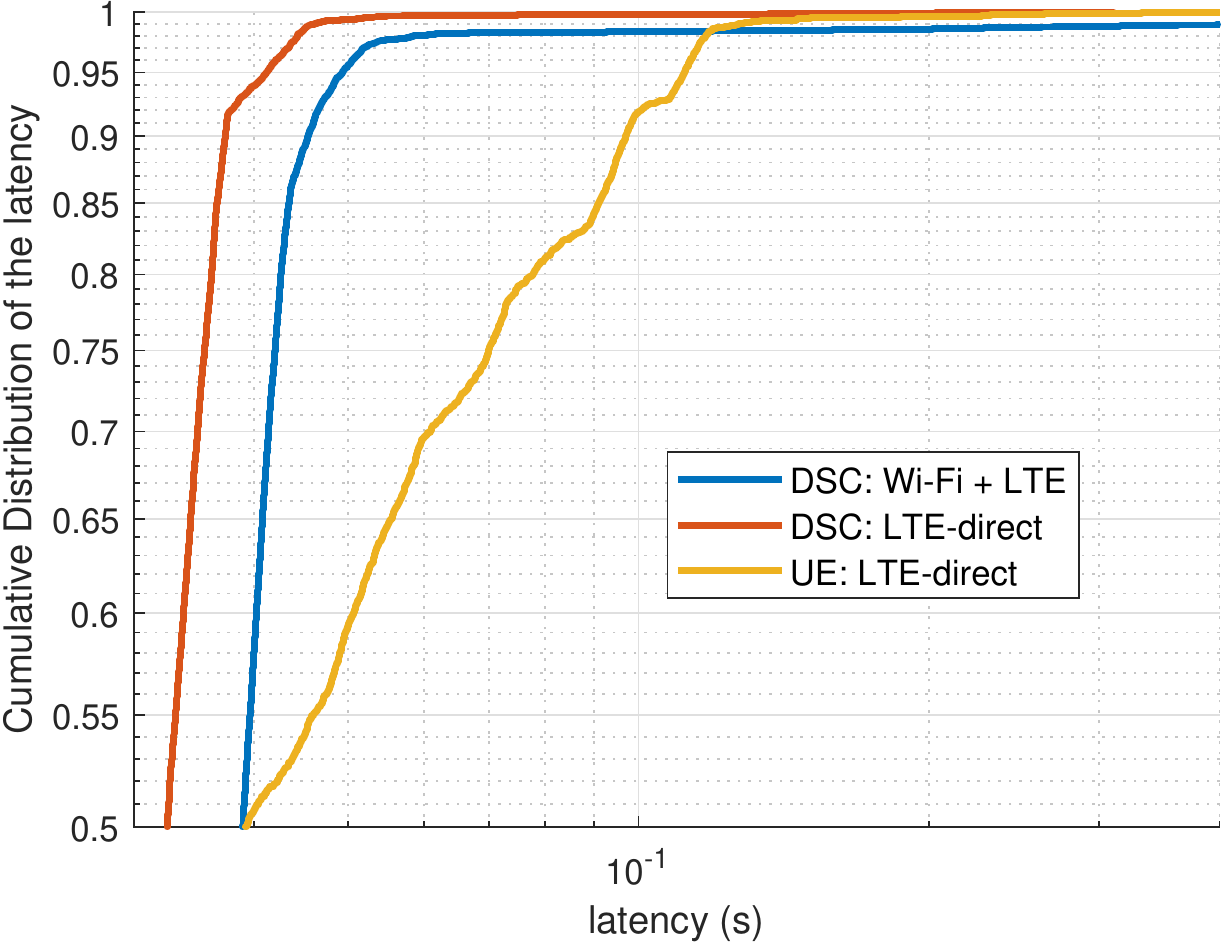}
\caption{Latency-Reliability plot of above average latency for each connection.
}
\label{fig:latrel}
\end{figure}

Even with this advantage of the UAV, the mean latency for the DSC system is nearly identical. Therefore, in Fig.~\ref{fig:latrel} we plot the empirical cumulative distribution function (ECDF) starting at the average latency, as taken from all runs. 
While the average latency is similar for the LTE-direct link and the DSC link, the variations are much higher for the former. In particular, it is much more likely to observe highly varying latencies between 40\,ms and 120\,ms when having a direct LTE connection. In other words, the LTE-direct link suffers from high jitter. This is not true for the DSC system that offers low jitter and a latency below 50\,ms with 0.95 probability. Unfortunately, due to the behavior of CSMA it is likely to observe a latency in the excess of 200\,ms, for 2\% of the time of operation of the DSC system. We note that this investigation is limited to generally good CSMA congestion conditions, and larger delays can be introduced when sharing the Wi-Fi carrier with other devices. Therefore, through the choice of location we also impact the wireless channel congestion for both the LTE and Wi-Fi spectrum. As such, the analysis presented in this work can be further enhanced by combining the provided measurements with congestion models, which is out of the scope of this paper.

\begin{figure}[t]
\centering
\includegraphics[width=0.92\columnwidth]{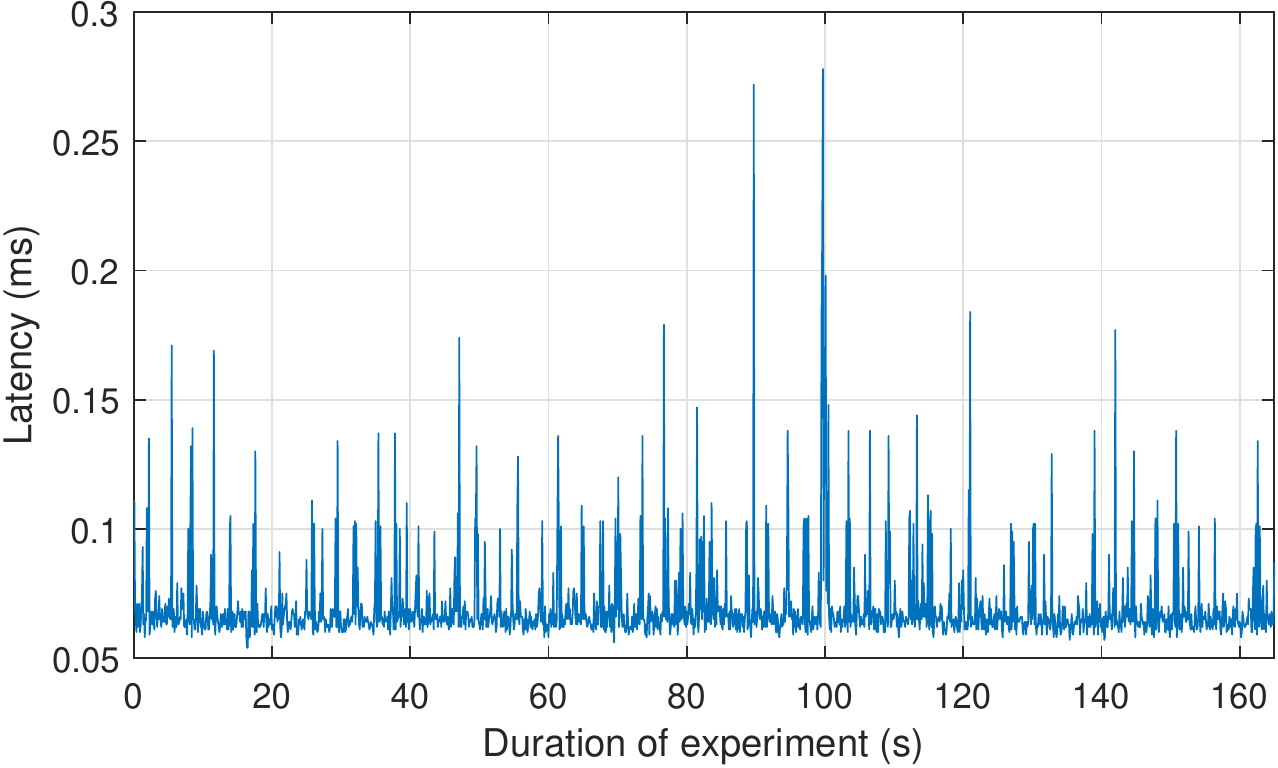}
\caption{Latency, in milliseconds, induced due to total time spent processing in DSC. 
}
\label{fig:proc}
\end{figure}
Finally, we observe the latencies in Fig.~\ref{fig:proc} that are the result of scheduling and processing of our computational system mounted on the drone.

\section{Conclusions}
\label{sec:conclusion}
In this paper, we conducted and elaborated an experimental analysis of the possibility of addressing latency concerns for ground based UE with DSCs. Initially, we provided a full sweep of measurements at different altitudes to inspect the impact of LoS and NLoS links. In the sweep we concluded that a DSC should operate at an altitude of 30\,m to provide a Wi-Fi hotspot for the UE below. In additional experiments, we compared the performance of an LTE-direct link -- between the UE and the BS -- to a DSC provided hotspot. The experimental analysis shows that the altitude of the DSC allows for reliably lower latency than an LTE-direct link from the ground. Moreover, the DSC system extended the coverage for an extra 6.4\% of the route. Even though this number is arbitrary and directly impacted by the choice of the flight and walking trajectories, it showcases that the DSC covers absolute edge cases of full link dropping. The drawbacks of the DSC system come due to the nature of Wi-Fi, as we occasionally observed latency spikes in the excess of 1\,s that are due to full outages. This removes the possibility of using the Wi-Fi fronthaul for ultra reliable remote control implementations that require low latency. Finally, this work allows us to segment the collective fronthaul, processing and backhaul latency of a complex DSC system in mixed LoS and NLoS scenarios, limited to outdoor users. In a future work we would like to observe the performance of a similar experimental setup with a 5G backhaul. Such an analysis would be imperative since next-generation communication equipment promises low latency even in NLoS environments. In those cases, upgrading the DSC fronthaul to the next-generation of 802.11be might finally achieve URLLC for outdoor ground users. Moreover, to evaluate the feasibility of such a system, we would also strive to reduce the complexity and weight of the mounted equipment in favor of increased drone air time. Such work can finally target the goal of URLLC for outdoor remote control applications in obstacle-dense environments, an evolution towards modular, fault-tolerable and reliable connectivity.

\section*{Acknowledgment}
The work was supported by the European Union's research and innovation programme under the Marie Sklodowska-Curie grant agreement No. 812991 ''PAINLESS'' within the Horizon 2020 Program. The experiments within the 5G Playground Carinthia are funded by the Carinthian Agency for Investment Promotion and Public Shareholding (BABEG). The 5G Playground is operated by BABEG and financed by means of the Austrian Federal Ministry for Climate Action, Environment, Energy, Mobility, Innovation and Technology (BMK) and the Carinthian provincial government.

\bibliography{main.bbl}


\begin{thebibliography}{18}


\ifx \showCODEN    \undefined \def \showCODEN     #1{\unskip}     \fi
\ifx \showDOI      \undefined \def \showDOI       #1{#1}\fi
\ifx \showISBNx    \undefined \def \showISBNx     #1{\unskip}     \fi
\ifx \showISBNxiii \undefined \def \showISBNxiii  #1{\unskip}     \fi
\ifx \showISSN     \undefined \def \showISSN      #1{\unskip}     \fi
\ifx \showLCCN     \undefined \def \showLCCN      #1{\unskip}     \fi
\ifx \shownote     \undefined \def \shownote      #1{#1}          \fi
\ifx \showarticletitle \undefined \def \showarticletitle #1{#1}   \fi
\ifx \showURL      \undefined \def \showURL       {\relax}        \fi
\providecommand\bibfield[2]{#2}
\providecommand\bibinfo[2]{#2}
\providecommand\natexlab[1]{#1}
\providecommand\showeprint[2][]{arXiv:#2}

\bibitem[\protect\citeauthoryear{Al-Hourani, Kandeepan, and
  Jamalipour}{Al-Hourani et~al\mbox{.}}{2014a}]%
        {atg}
\bibfield{author}{\bibinfo{person}{Akram Al-Hourani},
  \bibinfo{person}{Sithamparanathan Kandeepan}, {and} \bibinfo{person}{Abbas
  Jamalipour}.} \bibinfo{year}{2014}\natexlab{a}.
\newblock \showarticletitle{{Modeling Air-to-Ground Path Loss for Low Altitude
  Platforms in Urban Environments}}. In \bibinfo{booktitle}{\emph{Proc. of IEEE
  Global Communications Conference}}. \bibinfo{pages}{2898--2904}.
\newblock


\bibitem[\protect\citeauthoryear{Al-Hourani, Kandeepan, and Lardner}{Al-Hourani
  et~al\mbox{.}}{2014b}]%
        {optlap}
\bibfield{author}{\bibinfo{person}{Akram Al-Hourani},
  \bibinfo{person}{Sithamparanathan Kandeepan}, {and} \bibinfo{person}{Simon
  Lardner}.} \bibinfo{year}{2014}\natexlab{b}.
\newblock \showarticletitle{{Optimal LAP Altitude for Maximum Coverage}}.
\newblock \bibinfo{journal}{\emph{IEEE Wireless Communications Letters}}
  \bibinfo{volume}{3}, \bibinfo{number}{6} (\bibinfo{date}{Dec. ,}
  \bibinfo{year}{2014}), \bibinfo{pages}{569--572}.
\newblock


\bibitem[\protect\citeauthoryear{Athanasiadou, Batistatos, Zarbouti, and
  Tsoulos}{Athanasiadou et~al\mbox{.}}{2019}]%
        {athanasiadou2019lte}
\bibfield{author}{\bibinfo{person}{Georgia~E Athanasiadou},
  \bibinfo{person}{Michael~C Batistatos}, \bibinfo{person}{Dimitra~A Zarbouti},
  {and} \bibinfo{person}{George~V Tsoulos}.} \bibinfo{year}{2019}\natexlab{}.
\newblock \showarticletitle{LTE ground-to-air field measurements in the context
  of flying relays}.
\newblock \bibinfo{journal}{\emph{IEEE Wireless Communications}}
  \bibinfo{volume}{26}, \bibinfo{number}{1} (\bibinfo{year}{2019}),
  \bibinfo{pages}{12--17}.
\newblock


\bibitem[\protect\citeauthoryear{{Donevski} and {Nielsen}}{{Donevski} and
  {Nielsen}}{2020}]%
        {igoreucnc}
\bibfield{author}{\bibinfo{person}{I. {Donevski}} {and} \bibinfo{person}{J.~J.
  {Nielsen}}.} \bibinfo{year}{2020}\natexlab{}.
\newblock \showarticletitle{{Dynamic Standalone Drone-Mounted Small Cells}}. In
  \bibinfo{booktitle}{\emph{Proc. of European Conference on Networks and
  Communications (EuCNC)}}. \bibinfo{pages}{342--347}.
\newblock
\urldef\tempurl%
\url{https://doi.org/10.1109/EuCNC48522.2020.9200918}
\showDOI{\tempurl}


\bibitem[\protect\citeauthoryear{{Donevski}, {Nielsen}, and
  {Popovski}}{{Donevski} et~al\mbox{.}}{2021}]%
        {igorwcnc}
\bibfield{author}{\bibinfo{person}{I. {Donevski}}, \bibinfo{person}{J.~J.
  {Nielsen}}, {and} \bibinfo{person}{P. {Popovski}}.}
  \bibinfo{year}{2021}\natexlab{}.
\newblock \showarticletitle{{Standalone Deployment of a Dynamic Drone Cell for
  Wireless Connectivity of Two Services}}. In \bibinfo{booktitle}{\emph{Proc.
  of IEEE Wireless Communications and Networking Conference (WCNC)}}.
\newblock


\bibitem[\protect\citeauthoryear{Fakhreddine, Bettstetter, Hayat, Muzaffar, and
  Emini}{Fakhreddine et~al\mbox{.}}{2019}]%
        {fakhreddine2019handover}
\bibfield{author}{\bibinfo{person}{Aymen Fakhreddine},
  \bibinfo{person}{Christian Bettstetter}, \bibinfo{person}{Samira Hayat},
  \bibinfo{person}{Raheeb Muzaffar}, {and} \bibinfo{person}{Driton Emini}.}
  \bibinfo{year}{2019}\natexlab{}.
\newblock \showarticletitle{Handover challenges for cellular-connected drones}.
  In \bibinfo{booktitle}{\emph{Proceedings of the 5th Workshop on Micro Aerial
  Vehicle Networks, Systems, and Applications}}. \bibinfo{pages}{9--14}.
\newblock


\bibitem[\protect\citeauthoryear{{Fotouhi}, {Qiang}, {Ding}, {Hassan},
  {Giordano}, {Garcia-Rodriguez}, and {Yuan}}{{Fotouhi} et~al\mbox{.}}{2019}]%
        {mainsurvey}
\bibfield{author}{\bibinfo{person}{A. {Fotouhi}}, \bibinfo{person}{H. {Qiang}},
  \bibinfo{person}{M. {Ding}}, \bibinfo{person}{M. {Hassan}},
  \bibinfo{person}{L.~G. {Giordano}}, \bibinfo{person}{A. {Garcia-Rodriguez}},
  {and} \bibinfo{person}{J. {Yuan}}.} \bibinfo{year}{2019}\natexlab{}.
\newblock \showarticletitle{{Survey on UAV Cellular Communications: Practical
  Aspects, Standardization Advancements, Regulation, and Security Challenges}}.
\newblock \bibinfo{journal}{\emph{IEEE Communications Surveys Tutorials}}
  \bibinfo{volume}{21}, \bibinfo{number}{4} (\bibinfo{date}{Mar. ,}
  \bibinfo{year}{2019}), \bibinfo{pages}{3417--3442}.
\newblock
\urldef\tempurl%
\url{https://doi.org/10.1109/COMST.2019.2906228}
\showDOI{\tempurl}


\bibitem[\protect\citeauthoryear{Gangula, Esrafilian, Gesbert, Roux,
  Kaltenberger, and Knopp}{Gangula et~al\mbox{.}}{2018}]%
        {gangula2018flying}
\bibfield{author}{\bibinfo{person}{Rajeev Gangula}, \bibinfo{person}{Omid
  Esrafilian}, \bibinfo{person}{David Gesbert}, \bibinfo{person}{Cedric Roux},
  \bibinfo{person}{Florian Kaltenberger}, {and} \bibinfo{person}{Raymond
  Knopp}.} \bibinfo{year}{2018}\natexlab{}.
\newblock \showarticletitle{Flying rebots: First results on an autonomous
  UAV-based LTE relay using open airinterface}. In
  \bibinfo{booktitle}{\emph{2018 IEEE 19th International Workshop on Signal
  Processing Advances in Wireless Communications (SPAWC)}}. IEEE,
  \bibinfo{pages}{1--5}.
\newblock


\bibitem[\protect\citeauthoryear{Hayat, Bettstetter, Fakhreddine, Muzaffar, and
  Emini}{Hayat et~al\mbox{.}}{2019}]%
        {hayat2019experimental}
\bibfield{author}{\bibinfo{person}{Samira Hayat}, \bibinfo{person}{Christian
  Bettstetter}, \bibinfo{person}{Aymen Fakhreddine}, \bibinfo{person}{Raheeb
  Muzaffar}, {and} \bibinfo{person}{Driton Emini}.}
  \bibinfo{year}{2019}\natexlab{}.
\newblock \showarticletitle{An experimental evaluation of LTE-A throughput for
  drones}. In \bibinfo{booktitle}{\emph{Proceedings of the 5th Workshop on
  Micro Aerial Vehicle Networks, Systems, and Applications}}.
  \bibinfo{pages}{3--8}.
\newblock


\bibitem[\protect\citeauthoryear{Kovacs, Amorim, Nguyen, Wigard, and
  Mogensen}{Kovacs et~al\mbox{.}}{2017}]%
        {kovacs2017interference}
\bibfield{author}{\bibinfo{person}{Istvan Kovacs}, \bibinfo{person}{Rafhael
  Amorim}, \bibinfo{person}{Huan~C Nguyen}, \bibinfo{person}{Jeroen Wigard},
  {and} \bibinfo{person}{Preben Mogensen}.} \bibinfo{year}{2017}\natexlab{}.
\newblock \showarticletitle{Interference analysis for UAV connectivity over LTE
  using aerial radio measurements}. In \bibinfo{booktitle}{\emph{2017 IEEE 86th
  Vehicular Technology Conference (VTC-Fall)}}. IEEE, \bibinfo{pages}{1--6}.
\newblock


\bibitem[\protect\citeauthoryear{Li, Jiang, Sun, Cai, and Wen}{Li
  et~al\mbox{.}}{2016}]%
        {li2016development}
\bibfield{author}{\bibinfo{person}{Boyang Li}, \bibinfo{person}{Yifan Jiang},
  \bibinfo{person}{Jingxuan Sun}, \bibinfo{person}{Lingfeng Cai}, {and}
  \bibinfo{person}{Chih-Yung Wen}.} \bibinfo{year}{2016}\natexlab{}.
\newblock \showarticletitle{Development and testing of a two-UAV communication
  relay system}.
\newblock \bibinfo{journal}{\emph{Sensors}} \bibinfo{volume}{16},
  \bibinfo{number}{10} (\bibinfo{year}{2016}), \bibinfo{pages}{1696}.
\newblock


\bibitem[\protect\citeauthoryear{Lin, Wiren, Euler, Sadam,
  M{\"a}{\"a}tt{\"a}nen, Muruganathan, Gao, Wang, Kauppi, Zou,
  et~al\mbox{.}}{Lin et~al\mbox{.}}{2019}]%
        {lin2019mobile}
\bibfield{author}{\bibinfo{person}{Xingqin Lin}, \bibinfo{person}{Richard
  Wiren}, \bibinfo{person}{Sebastian Euler}, \bibinfo{person}{Arvi Sadam},
  \bibinfo{person}{Helka-Liina M{\"a}{\"a}tt{\"a}nen}, \bibinfo{person}{Siva
  Muruganathan}, \bibinfo{person}{Shiwei Gao}, \bibinfo{person}{Y-P~Eric Wang},
  \bibinfo{person}{Juhani Kauppi}, \bibinfo{person}{Zhenhua Zou},
  {et~al\mbox{.}}} \bibinfo{year}{2019}\natexlab{}.
\newblock \showarticletitle{Mobile network-connected drones: Field trials,
  simulations, and design insights}.
\newblock \bibinfo{journal}{\emph{IEEE Vehicular Technology Magazine}}
  \bibinfo{volume}{14}, \bibinfo{number}{3} (\bibinfo{year}{2019}),
  \bibinfo{pages}{115--125}.
\newblock


\bibitem[\protect\citeauthoryear{{Mozaffari}, {Saad}, {Bennis}, {Nam}, and
  {Debbah}}{{Mozaffari} et~al\mbox{.}}{2019}]%
        {tutorial}
\bibfield{author}{\bibinfo{person}{M. {Mozaffari}}, \bibinfo{person}{W.
  {Saad}}, \bibinfo{person}{M. {Bennis}}, \bibinfo{person}{Y. {Nam}}, {and}
  \bibinfo{person}{M. {Debbah}}.} \bibinfo{year}{2019}\natexlab{}.
\newblock \showarticletitle{{A Tutorial on UAVs for Wireless Networks:
  Applications, Challenges, and Open Problems}}.
\newblock \bibinfo{journal}{\emph{IEEE Communications Surveys Tutorials}}
  \bibinfo{volume}{21}, \bibinfo{number}{3} (\bibinfo{date}{Mar. ,}
  \bibinfo{year}{2019}), \bibinfo{pages}{2334--2360}.
\newblock
\urldef\tempurl%
\url{https://doi.org/10.1109/COMST.2019.2902862}
\showDOI{\tempurl}


\bibitem[\protect\citeauthoryear{Muzaffar, Raffelsberger, Fakhreddine, Luque,
  Emini, and Bettstetter}{Muzaffar et~al\mbox{.}}{2020}]%
        {muzaffar2020first}
\bibfield{author}{\bibinfo{person}{Raheeb Muzaffar}, \bibinfo{person}{Christian
  Raffelsberger}, \bibinfo{person}{Aymen Fakhreddine},
  \bibinfo{person}{Jos{\'e}~L{\'o}pez Luque}, \bibinfo{person}{Driton Emini},
  {and} \bibinfo{person}{Christian Bettstetter}.}
  \bibinfo{year}{2020}\natexlab{}.
\newblock \showarticletitle{First experiments with a 5G-Connected drone}. In
  \bibinfo{booktitle}{\emph{Proceedings of the 6th ACM Workshop on Micro Aerial
  Vehicle Networks, Systems, and Applications}}. \bibinfo{pages}{1--5}.
\newblock


\bibitem[\protect\citeauthoryear{Nguyen, Amorim, Wigard, Kov{\'a}cs,
  S{\o}rensen, and Mogensen}{Nguyen et~al\mbox{.}}{2018}]%
        {nguyen2018ensure}
\bibfield{author}{\bibinfo{person}{Huan~Cong Nguyen}, \bibinfo{person}{Rafhael
  Amorim}, \bibinfo{person}{Jeroen Wigard}, \bibinfo{person}{Istv{\'a}n~Z
  Kov{\'a}cs}, \bibinfo{person}{Troels~B S{\o}rensen}, {and}
  \bibinfo{person}{Preben~E Mogensen}.} \bibinfo{year}{2018}\natexlab{}.
\newblock \showarticletitle{How to ensure reliable connectivity for aerial
  vehicles over cellular networks}.
\newblock \bibinfo{journal}{\emph{IEEE Access}}  \bibinfo{volume}{6}
  (\bibinfo{year}{2018}), \bibinfo{pages}{12304--12317}.
\newblock


\bibitem[\protect\citeauthoryear{Qiu, Chu, Calvo-Ramirez, Briso, and Yin}{Qiu
  et~al\mbox{.}}{2017}]%
        {qiu2017low}
\bibfield{author}{\bibinfo{person}{Zhihong Qiu}, \bibinfo{person}{Xi Chu},
  \bibinfo{person}{Cesar Calvo-Ramirez}, \bibinfo{person}{C{\'e}sar Briso},
  {and} \bibinfo{person}{Xuefeng Yin}.} \bibinfo{year}{2017}\natexlab{}.
\newblock \showarticletitle{Low altitude UAV air-to-ground channel measurement
  and modeling in semiurban environments}.
\newblock \bibinfo{journal}{\emph{Wireless Communications and Mobile
  Computing}}  \bibinfo{volume}{2017} (\bibinfo{year}{2017}).
\newblock


\bibitem[\protect\citeauthoryear{Raffelsberger, Muzaffar, and
  Bettstetter}{Raffelsberger et~al\mbox{.}}{2019}]%
        {cdmt}
\bibfield{author}{\bibinfo{person}{Christian Raffelsberger},
  \bibinfo{person}{Raheeb Muzaffar}, {and} \bibinfo{person}{Christian
  Bettstetter}.} \bibinfo{year}{2019}\natexlab{}.
\newblock \showarticletitle{A performance evaluation tool for drone
  communications in 4G cellular networks}. In \bibinfo{booktitle}{\emph{2019
  16th International Symposium on Wireless Communication Systems (ISWCS)}}.
  IEEE, \bibinfo{pages}{218--221}.
\newblock


\bibitem[\protect\citeauthoryear{Van~der Bergh, Chiumento, and Pollin}{Van~der
  Bergh et~al\mbox{.}}{2016}]%
        {van2016lte}
\bibfield{author}{\bibinfo{person}{Bertold Van~der Bergh},
  \bibinfo{person}{Alessandro Chiumento}, {and} \bibinfo{person}{Sofie
  Pollin}.} \bibinfo{year}{2016}\natexlab{}.
\newblock \showarticletitle{LTE in the sky: Trading off propagation benefits
  with interference costs for aerial nodes}.
\newblock \bibinfo{journal}{\emph{IEEE Communications Magazine}}
  \bibinfo{volume}{54}, \bibinfo{number}{5} (\bibinfo{year}{2016}),
  \bibinfo{pages}{44--50}.
\newblock


\end{thebibliography}
\end{document}